\documentclass[twocolumn]{aastex63}
\usepackage{graphicx}	
\usepackage{amsmath}	
\usepackage{amssymb}	
\usepackage{ulem}
\usepackage{tabularx,booktabs}

\received{December xx, 2019}
\revised{December xx, 2019}
\accepted{January 02, 2020}
\submitjournal{\apjl}
\shorttitle{Temporal properties of MAXI J1820+070}
\shortauthors{Mudambi S. P. et al.}
\begin{document}
\title{Unveiling the temporal properties of MAXI J1820+070 through {\it{AstroSat}} observations}
\correspondingauthor{Shivappa B. Gudennavar}
\email{shivappa.b.gudennavar@christuniversity.in, rmisra@iucaa.in,
sneha.m@res.christuniversity.in}
\author[0000-0003-0076-4441]{Sneha Prakash Mudambi}
\affiliation{Department of Physics and Electronics, CHRIST (Deemed to be University), Bangalore Central Campus, Bengaluru-560029, India}
\author[0000-0002-0360-1851]{Bari Maqbool}
\affiliation{Inter-University Centre for Astronomy and Astrophysics, Ganeshkind, Pune-411007, India}
\author{Ranjeev Misra}
\affiliation{Inter-University Centre for Astronomy and Astrophysics, Ganeshkind, Pune-411007, India}
\author[0000-0003-0507-5709]{Sabhya Hebbar}
\affiliation{Department of Physics and Electronics, CHRIST (Deemed to be University), Bangalore Central Campus, Bengaluru-560029, India}
\author{J. S. Yadav}
\affiliation{Department of Physics, Indian Institute of Technology, Kanpur-208016, India}
\author[0000-0002-9019-9441]{Shivappa B. Gudennavar}
\affiliation{Department of Physics and Electronics, CHRIST (Deemed to be University), Bangalore Central Campus, Bengaluru-560029, India}
\author[0000-0003-1234-0662]{Bubbly S. G.}
\affiliation{Department of Physics and Electronics, CHRIST (Deemed to be University), Bangalore Central Campus, Bengaluru-560029, India}
\begin{abstract}
\noindent
We present here the results of the first broadband simultaneous spectral and temporal studies of the newly detected black hole binary MAXI J1820+070 as seen by SXT and LAXPC on-board {\it{AstroSat}}. The observed combined spectra in the energy range 0.7$-$80~keV were well modeled using disk blackbody emission, thermal Comptonization and a reflection component. The spectral analysis revealed that the source was in its hard spectral state (\textit{{$\Gamma$}}~=~1.61) with a cool disk ($kT_{in}$~=~0.22~keV). We report the energy dependent time-lag and root mean squared (rms) variability at different frequencies in the energy range 3$-$80~keV using LAXPC data. We also modeled the flux variability using a single zone stochastic propagation model to quantify the observed energy dependence of time-lag and fractional rms variability and then compared the results with that of Cygnus~X-1. Additionally, we confirm the detection of a quasi-periodic oscillation with the centroid frequency at 47.7~mHz.
\end{abstract}
\section{Introduction}
\noindent
Accreting black hole X-ray binaries (BHXBs) in outburst exhibit random short-term variability in their flux \citep{Van89} which may arise due to the perturbations occurring at different radii of the accretion disk propagating inwards \citep{Lyu97}. These perturbations cause variations to mass accretion rate at the inner regions of accretion disk on different time-scales \citep{Spr87,Nar94,Abr95,Che95}. The X-ray variability in BHXBs  are well represented by their power density spectra (PDS) which exhibit  systematic  changes  throughout  the course  of an outburst with remarkable similarities among themselves, thereby suggesting a common underlying physical  phenomenon \cite[and references therein]{Bel10}. The PDS of most BHXBs are characterized by broadband continuum noise like features and sometimes narrow peak features called quasi-periodic oscillations (QPOs). The exact mechanism of origin of QPOs is still an open question but the  origin of broadband noise could be due to the inward propagation and coupling of perturbations occurring throughout the accretion disk resulting in flux variations. This scenario best explains the observed linearity of the rms-flux relationship in galactic black holes \citep{Gle04} and other type of X-ray sources \citep{Utt01,Gas04}. Furthermore, \citet{Hei12} showed that the rms-flux relationship of broadband noise is an universal feature of all accreting BHXBs, independent of their spectral state.
\\[4pt]
Over the past two decades, there have been several efforts to develop a unified propagation fluctuation model to explain and predict the energy and frequency dependent fractional rms and time-lag \citep{Ing11, Ing12, Ing13, Rap16, Rap17, Axe18}. The motivation being that such efforts could provide the necessary tools to probe the geometry of the system within this regime and possibly explain the exact mechanism of origin of QPOs \citep{Bot99, Mis00, Kot01}.
\\[4pt]
\citet{Maq19} proposed and validated one such model by comparing its predictions with observed data of Cygnus~X-1 from \textit{AstroSat}. The model invokes a simple geometry of standard truncated disk with a hot inner region \citep{Esi97} and assumes that hard X-ray component originates from this hot inner region by a single temperature thermal Comptonization process. The model considers variation of the temperature of the inner radius of the truncated disk and  that of the hot inner flow with a frequency dependent time-lag between them. Data from Soft X-ray Telescope (SXT) and Large Area X-ray Proportional Counter (LAXPC) on-board \textit{AstroSat} play a crucial role for validating the model as it provides unprecedented spectro-timing information with broadband coverage \citep{Mis17}. The model successfully explained the energy dependent rms and time-lags in Cygnus~X-1. Nonetheless, the results obtained are for a persistent BHXB in its hard state. It is unclear how the results would change for a transient BHXB in the same state thereby making it necessary to test and validate the model on different types of X-ray binaries.
\\[4pt]
In view of this, we considered \textit{AstroSat}'s data of the newly discovered transient, MAXI J1820+070 to validate the model proposed by \citet{Maq19}. MAXI J1820+070, a galactic black hole X-ray transient, is one among the brightest X-ray novae observed till date \citep{Cor16}. The source was first discovered in optical on 6$^{th}$ March 2018 by \textit{ASSAS-SN} project \citep{Tuc18} and later in X-rays on 11$^{th}$ March 2018 by \textit{MAXI} \citep{Kaw18}. Subsequent multi-wavelength observations revealed that MAXI J1820+070 is a BHXB system \citep{Bag18,Utt18}. The outburst cycle which lasted for almost a year showed rapid, frequent, alternating transitions between hard$-$soft spectral states \citep{Rus19}. \citet{Shi19} analyzed the data from \textit{MAXI} and \textit{SWIFT} spanning the entire outburst and reported the similarities between two re-brightening events that occurred in March and June 2018. Interestingly, these re-brightening events were peaked at comparable X-ray luminosity in 2$-$20~keV despite showing hard$-$soft spectral state transition only in one of them (June 2018).
\\[4pt]
\citet{Gan19} reported the source distance to be 3.46$^{+2.18}_{-1.03}$~kpc using DR2 data of \textit{Gaia}.~\citet{Utt18} found the Galactic extinction to be 1.5$\times10^{21}$~cm$^{-2}$ by analyzing the soft X-ray data from \textit{NICER} during the rising phase of outburst. \citet{Kar19} reported high frequency reverberation time-lags (soft lags) between the energy bands 0.1$-$1~keV and 1$-$10~keV also using the \textit{NICER} data, which was seen for the first time for a BHXB. \citet{Bha19} constrained the physical parameters of the black hole utilizing the data from \textit{NuSTAR} and \textit{SWIFT/XRT}. The combined spectral fit revealed the presence of weak disk blackbody emission and dominant thermal Comptonization along with relativistic reflection fraction. Furthermore, they estimated the source inclination angle to be $\sim30^{\circ}$ and the inner disk radius to be $\sim$~4.2 times the radius of the innermost stable circular orbit.
\\[4pt]
The proximity of the source, high flux rate of $\sim$~10$^{-8}$~erg cm$^{-2}$s$^{-1}$ and low galactic extinction make MAXI J1820+070 an ideal candidate to test stochastic propagation model proposed by \citet{Maq19}. In this study, we report the first simultaneous broadband spectro$-$timing results of MAXI J1820+070 using data from SXT and LAXPC on-board \textit{AstroSat}. We also report the results of the stochastic propagation model fit and their comparison with the that of Cygnus~X-1.
\section{Data reduction and analysis}
\subsection{Data reduction}
\noindent
We have analyzed $\sim$~93~ks simultaneous data from SXT and LAXPC on-board \textit{AstroSat} spanning over a period of 2 days starting from 30$^{th}$ March 2018. Observed data consisted of 15 continuous segments corresponding to 15 individual orbits of the satellite. Level 1 photon counting mode data of SXT was processed through the official SXT pipeline  AS1SXTLevel2-1.4b\footnote{http://www.tifr.res.in/~astrosat$\_$sxt/sxtpipeline.html} to produce Level 2 files.
\\[4pt]
The HEASoft version 6.24 tool XSELECT was used to extract spectra and lightcurve between the source regions of 4\arcmin~(inner radius) and 16\arcmin~(outer radius). These regions were chosen to account for pile up effect in the charged coupled device (CCD) due to high flux rate ($\sim$~1~Crab) of the source in SXT energy range (0.3$-$8.0~keV). The response matrix file (RMF) `sxt$\_$pc$\_$mat$\_$g0to12.rmf', standard background spectrum `SkyBkg$\_$comb$\_$EL3p5$\_$Cl$\_$Rd16p0$\_$v01.pha' and an ancillary response file (ARF) appropriate for source location on the CCD created using SXT ARF generation tools \footnote{http://www.tifr.res.in/~astrosat$\_$sxt/dataanalysis.html} were used for the analysis. 
\\[4pt]
Each of the fifteen images were visually examined to ensure that drift corrections of the satellite were applied and a single point source was seen. Following this, we used SXT event merger tool\footnote{http://www.tifr.res.in/~astrosat$\_$sxt/dataanalysis.html} to merge all fifteen individual Level 2 files into one single merged event file. The merged file was later used for spectral analysis.
\\[4pt]
The official LAXPC software\footnote{http://astrosat-ssc.iucaa.in/?q=laxpcData} was used to process Level 1 event mode data to obtain Level 2 files. The sub-routines of LAXPC software \citep{Ant17} were used to generate good time interval file, total spectrum, RMF and background spectrum for proportional counters 10, 20 and 30 respectively. However, data from LAXPC10 and LAXPC30 were not used for this work as the POC team reported abnormal gain change in LAXPC10 on 28$^{th}$ March 2018 and LAXPC30 was not operational during this observation.
\subsection{Spectral analysis}
\noindent
We have performed a combined spectral fitting of SXT and LAXPC20 spectra using XSPEC 12.9.1p in the energy range 0.7$-$80~keV. Lower energies ($<$~0.7~keV) were not considered due to the uncertainties in the effective area and response of CCD. A 3$\%$ systematic error was incorporated during analysis to account for uncertainties in response calibration. Background uncertainty was taken to be 3$\%$. Gain correction was applied to the SXT data using the XSPEC command {\it gain fit} with slope unity and the best fit offset value was found to be 23~eV. Relative normalization was allowed to vary between SXT and LAXPC data.
\\[4pt]
The source spectrum was found to be dominated by thermal Comptonization component along with disk emission and reflection component (Figure~\ref{powspec}, Left panel). We used XSPEC models {\it nthcomp} \citep{Zyc99}, {\it diskbb} \citep{Mit84,Mak86} and {\it ireflect} \citep{Mag95} to fit the thermal Comptonization, disk emission and reflection component respectively. We also added model {\it tbabs} \citep{Wil00} to account for interstellar absorption. Interstellar hydrogen column density ($N_{H}$) was frozen at $0.15\times10^{22}$~cm$^{-2}$ \citep{Utt18}. Disk emission was considered to be input seed photons for Comptonization and thus the parameters $T_{in}$ and {\textit{nthcomp}} temperature ($kT_{bb}$) were tied together and treated as a single parameter during the fitting. For the reflection model, the abundance was fixed to solar values while the inclination was taken to be $\sim$~$30^{\circ}$ \citep{Bha19}. The temperature of the reflector was tied to the inner disk temperature. Electron temperature of the Comptonizing cloud ($kT_{e}$) was not constrained by the data and hence was fixed at a fiduciary value of 100~keV. Disk ionization parameter was frozen at 10~ergcms$^{-1}$. Best fit spectral parameters are listed in Table~\ref{specpar}. Spectral index ($\Gamma$~=~1.61$\pm0.01$) obtained indicated that the source was in hard spectral state \citep{Tit05}. Disk temperature was found to be 0.22$\pm0.01$~keV suggesting a cool disk truncated at a large distance ($\sim$~526~km) \citep{Kub98}. Spectra also showed the presence of a Compton hump around 30~keV indicating disk reflection. Note that the parameters obtained here are an approximate due to the large systematic error~(3$\%$) considered while fitting SXT and LAXPC spectra, which did not allow for more precise spectral modeling. We also tried using few other reflection models like \textit{reflionx} to further constrain the reflection parameters. However, we did not find any significant improvement in the fit or obtain a better estimate of the parameters. 
\subsection{Temporal analysis}
\begin{figure*}
\centering
\includegraphics[width=8cm,height=5cm]{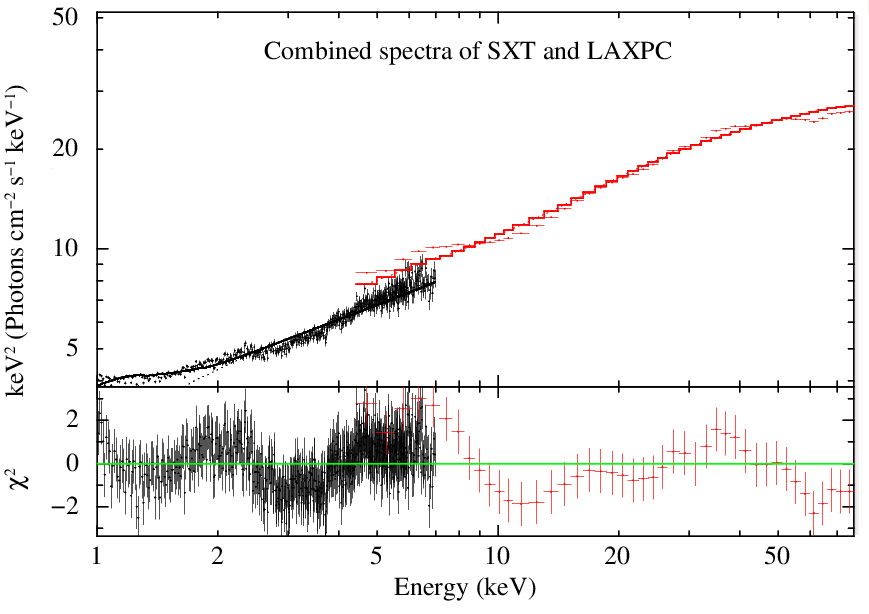}
\includegraphics[width=8cm,height=4.95cm]{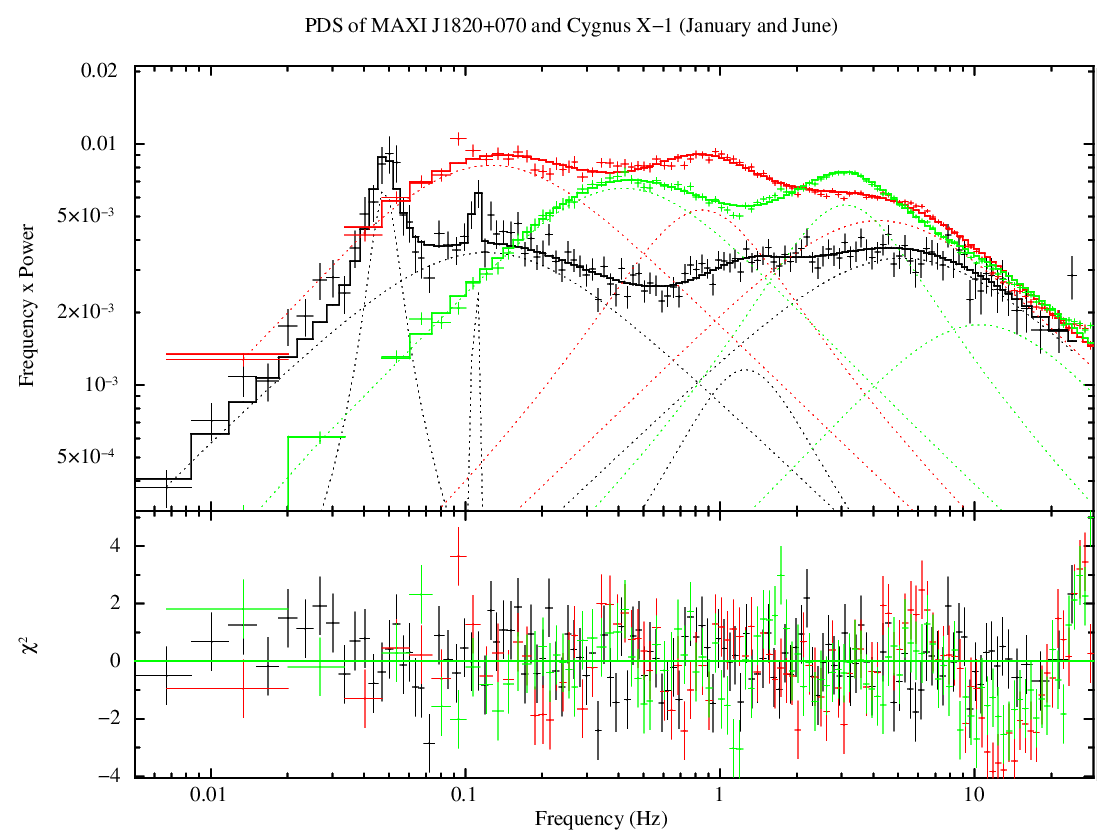}
\caption{Fitted SXT and LAXPC spectra and residuals for all the 15 orbits (Left panel). Comparative PDS of MAXI J1820+070 (Black color) and Cygnus~X-1 January (Green color) and June (Red color) 2016 in the 3$-$80~keV energy band. PDS of the source is fitted with five Lorentzian components of which the two narrow peaks represent a QPO at 47.7~mHz and a weak oscillation at 109.4~mHz respectively. PDS of Cygnus~X-1 is fitted with 2 and 3 Lorentzian components for January and June respectively (Right panel)}.
\label{powspec}
\end{figure*}

\begin{figure*}
\centering
\includegraphics[width=18cm,height=9cm]{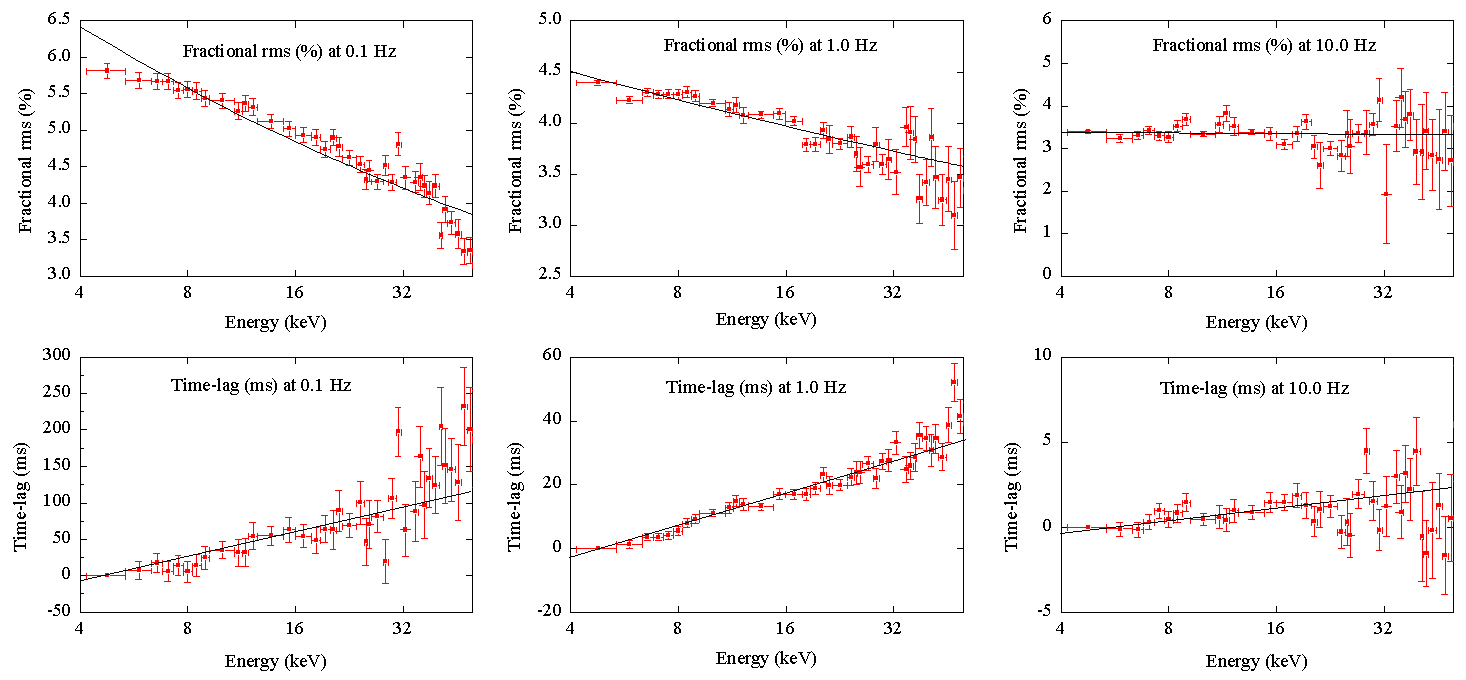}
\caption{Empirical fit to the observed energy dependent fractional rms variability (Top panel) and time-lag (Bottom panel) at 3 fundamental frequencies 0.1, 1.0 and 10.0~Hz as a function of energy.
\label{empfit}}
\end{figure*}

\noindent
PDS was generated in the energy range 3$-$80~keV using LAXPC20 data by averaging segments of 282.64~s with 0.0045~s time resolution. It was then binned logarithmically in frequency to obtain a PDS shown in the Right panel of Figure~\ref{powspec} in the frequency range 0.004$-$30 Hz. The spectrum shows a prominent QPO at 47.7~mHz and three broadened noise humps, which can be represented by Lorentzians.
There is also a weak feature at 109.4~mHz and modeling this component with a Lorentzian resulted in a decrease of $\chi^2$, i.e. $\Delta~\chi^2$~=~12 for three additional degrees of freedom (dof). The final $\chi^2$/(dof) was found to be 123/114 after taking into account 2\% systematic error. The best fit parameters are listed in Table~\ref{specpar}. 
\\[4pt]
Following this, we studied energy dependent temporal behavior by adopting the methodology discussed in \citet{Maq19}. The complete event mode data available from LAXPC for computation of fractional rms and time-lag for a large number of finer energy bins integrated over a certain frequency range allows one to compare the results directly with models. 
\begin{table*}
\caption{Spectral and PDS parameters}
\begin{flushleft}
\begin{tabular}{llcclll}
\toprule
\textbf{Spectral parameters}&&&&&&\\
\hline
Model parameters&rel$\_$refl&$\Gamma$&$N_{comp}$&$kT_{in}$&$N_{disk}$ & $\chi{^2}$/dof \\
(Description)&(Relativistic&(Asymptotic& (Normalization&(Temperature&(Normalization\\
&reflection)&power-law)&factor)&at inner disk&factor)\\
&&&&radius) keV&$\times 10^{5}$&\\
\midrule 
Best fit value&0.17$^{+0.04}_{-0.05}$&1.61$^{+0.01}_{-0.01}$&3.42$^{+0.05}_{-0.06}$&0.22$^{+0.01}_{-0.01}$&2.4$^{+0.6}_{-0.6}$&696/578\\
&&&&&& \\
\toprule 
\textbf{PDS parameters}&&&&&&\\
\midrule
Feature & QPO & Weak & Noise hump 1 & Noise hump 2 & Noise hump 3 & $\chi^{2}$/dof \\
& &oscillation &&&\\
\midrule
Centroid frequency (Hz) & 47.7$^{+1.6}_{-2.0}$ $\times10^{-3}$ & 109.4$^{+2.8}_{-1.2}$ $\times10^{-3}$& $0^{*}$ &$0^{*}$&1.04$^{+0.13}_{-0.20}$ \\
\\
Width (Hz) & 11.9$^{+5.0}_{-3.6}$ $\times10^{-3}$ & $<1.7 \times10^{-3}$ & 10.7$^{+2.3}_{-1.3}$ & 0.24$^{+0.02}_{-0.02}$ &1.6$^{+0.7}_{-0.6}$ & 123/114 \\\\
Norm $\times10^{-3}$ & 2.5$^{+0.6}_{-0.6}$ & 0.3$^{+0.2}_{-0.1}$ & 10.3$^{+0.8}_{-1.1}$ & 11.1$^{+0.6}_{-0.6}$ &2.1$^{+1.9}_{-0.9}$ \\
\bottomrule
\end{tabular}
\vspace{1pt}
$^{*}$Parameter frozen during fitting
\end{flushleft}
\label{specpar}
\end{table*}
\\[4pt]
We studied energy dependent fractional rms and time-lag in the energy range 3$-$80 keV for a range of frequencies using the data from LAXPC20. \citet{Now99} and \citet{Miy88} have shown that the energy dependence of time-lag is logarithmic in nature. Hence following the procedure discussed in \citet{Maq19} the observed time-lag and fractional rms were fit empirically using equations (\ref{eq1}) and (\ref{eq2}) respectively.
\begin{equation}
\delta t (E,f) = T_d (f)  \times log (\frac{E}{E_{ref}})  
\label{eq1}
\end{equation}
where $E$ is the energy (keV), $f$ is the frequency (Hz) under consideration, $E_{ref}$ = 4.76~keV is the reference energy and $T_d(f)$ is a constant. 
\begin{equation}
  {\rm F}(E,f) = A(f) \times E^{p(f)}
  \label{eq2}
\end{equation}
where $A(f)$ and $p(f)$ are constants. We have computed time-lag with respect to the reference energy band 4.15$-$5.37~keV for a wide range of frequencies. Figure~\ref{empfit} shows the energy dependent time-lag and fractional rms for three representative frequencies of 0.1, 1.0 and 10.0~Hz. The rms decreases with energy depending on the frequency whereas the time-lag increases with energy implying that they are hard lags where the hard energy photons lag the softer ones. Figure~\ref{modelfit} (Top panel) shows variation of the best fit parameters $A(f)$, $T_{d}(f)$ and $p(f)$ with frequency. 
\\[4pt]
Although Figure~\ref{modelfit} (Top panel) gives an idea about temporal behavior of the system in frequency and energy realm, they neither provide an interpretation of physical behavior of the system nor its association to the time-averaged PDS. Hence to make such an association, we have modeled the observed fractional rms and time-lag using the stochastic propagation model developed by \citet{Maq19}. We present this result along with a comparison of the results of Cygnus~X-1 in the following sections.
\subsection{Stochastic propagation model}
\begin{figure*}
\centering
\includegraphics[width=18cm,height=9cm]{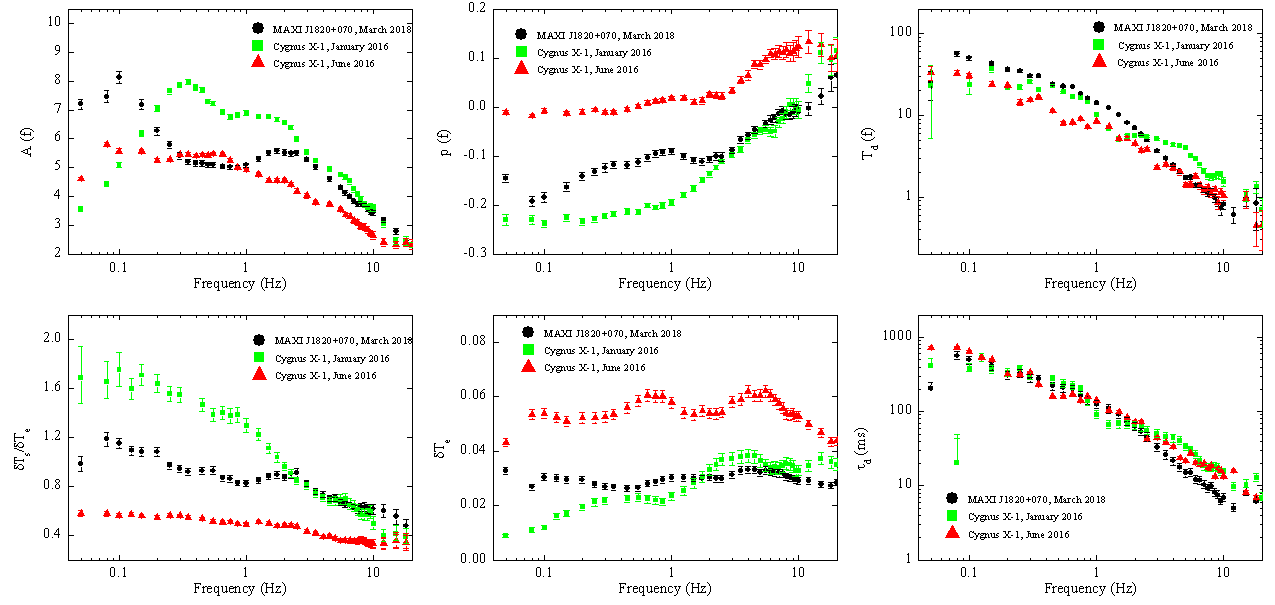}
\caption{Comparison of the empirical fit parameters (Top panel) and the model fit parameters (Bottom panel) for MAXI J1820+070 and Cygnus~X-1 (January and June 2016) as a function of frequency.}
\label{modelfit}
\end{figure*}

\begin{figure}
\centering
\includegraphics[width=8cm,height=5cm]{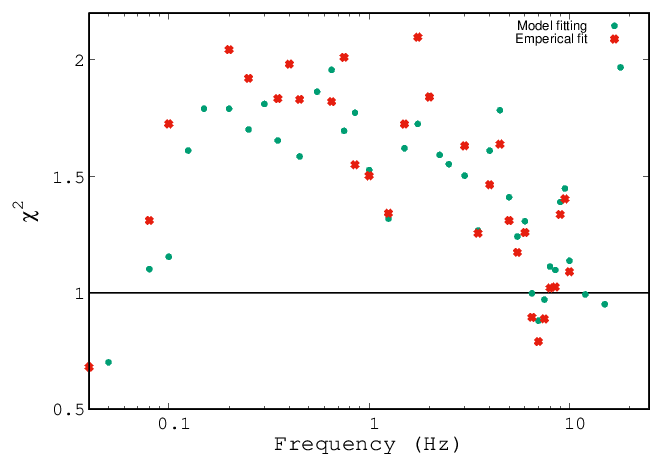}
\caption{Reduced $\chi{^2}$ values from the empirical and model fits plotted as a function of frequency.}
\label{model_var_fig}
\end{figure}

\noindent
The onset of this millennium has seen several ongoing efforts in the development of stochastic propagation models to explain energy dependent continuum variability observed in most X-ray binaries. Most of these models are built upon the fundamental idea that the perturbations originate at the outer annuli of the disk and propagate towards the central object \citep{Kot01,Ing11,Ing12,Ing13,Rap16,Rap17,Mah18,Axe18}. Primarily, these models are aimed at making quantitative predictions of PDS, energy dependent time-lags and rms variability. Time-lags are associated with perturbation propagation time (of the order of viscous time-scales) and prognosis of such studies could provide in-sights about geometry of the system in this scenario \citep{Bot99, Mis00, Kot01}.
\\[4pt]
\citet{Kot01} considered radial profile of the emissivity index and explained the frequency dependency of time-lag. \citet{Rap17} and \citet{Axe18} linked the production of hard X-rays to existence of several different Comptonization regions and the origin of photons of various energies with different regions of the accretion disk. However, it is difficult to comprehend the existence of such multi Comptonization regions, if high energy photons are indeed produced in the corona having optical depth $\sim$~1. Most importantly, this raises questions about the radial profile of emissivity index. On the other hand, \citet{Mis00} explained the frequency dependent time-lag by attributing the origin of hard X-ray photons to an optically thick disk with a rapidly varying temperature profile.
\\[4pt]
Recent works carried out by \citet{Cow14}, \citet{Hog16} and \citet{Ahm18} have shown that for standard optically thick and geometrically thin disks the propagation time-scales are frequency dependent and are different from the viscous ones. \citet{Maq19} quantified this behavior for hot, optically thin, geometrically thick flows considering the propagation time from the transition region onwards. They also considered the Comptonized spectrum to arise from a single Comptonizing zone as opposed to multi Comptonizing zones assumed in previous works. They showed that the observed frequency dependent time-lag between various energy bands may be caused due to underlying time-lag between seed photon fluctuations and subsequent variation of heating rate of the hot inner flow.
\\[4pt]
\noindent
The primary goal here is to quantify fractional rms and time-lag measured by LAXPC as described in \S 2.3 in terms of the model discussed in \citet{Maq19}. The model is characterized by three parameters photon index ($\Gamma$), electron temperature ($T_{e}$) and blackbody temperature ($T_{bb}$), which are obtained from the spectral fitting described in \S 2.2. Interestingly, the model uses the same number of parameters as in the empirical fitting described in \S 2.3. We were able to quantitatively explain the observed variability using only three parameters (the normalized amplitudes of temperature of the truncated disk ($\delta T_{s}$), hot inner flow ($\delta T_{e}$) and the time-lag between them ($\tau_{D}$)). The results of the model fitting are shown in Figure~\ref{modelfit} (Bottom panel), where $\delta T_{e}$, $\tau_{D}$ and the ratio $\delta T_{s}/\delta T_{e}$ are plotted against frequency. The variation of $\delta T_e$ shows three broadened humps which reflect the features exhibited in the PDS described in \S 2.3. Time delay or time-lag ($\tau_{D}$) between the variation of disk temperature and that of the Comptonizing cloud is a function of frequency. At frequencies $<$2~Hz the lags are of the order of 100~ms, where as at frequencies $>$2~Hz one sees a time-delay of the order of 10~ms. The ratio $\delta T_{s}/\delta T_{e}$ represents the attenuation of propagation from the disk to corona. This ratio is about 1.2 at lower frequencies ($<$0.2~Hz) and roughly flattens out until 2~Hz, just before dropping slightly at higher frequencies ($>$2~Hz). The results of the modeling showed good agreement with the empirical fit. The reduced $\chi{^2}$ in both cases (empirical and model fit) were around 2 and showed similar trend (Figure~\ref{model_var_fig}) thereby making the empirical functions defined in \S 2.3 adequate.
\subsection{Comparison with Cygnus~X-1}
\noindent
Most BHXBs in their hard state show similar spectro-timing characteristics \citep{Utt14}. Cygnus~X-1 being the most studied X-ray source by generations of X-ray missions spends most of its time in hard state \citep{Zha97}. 
The spectro-timing features of Cygnus~X-1 are relatively well known as compared to other BHXBs thus making it suitable for comparative studies.
\\[4pt]
\citet{Maq19} studied six \textit{AstroSat} observations of Cygnus~X-1 in its hard state, spread out between January and October 2016. Spectral analysis of these observations revealed a cooler disk ($\sim$~0.46~keV) with large \textit{diskbb} normalization ($\sim$~2400) for January 2016 as compared to rest of the months. Moreover, the PDS of January was found to have only two broadened noise humps whereas rest of the months showed three humps (Figure~\ref{powspec}, Right panel). In this work, we have compared our temporal analysis results with January and June 2016 data of Cygnus~X-1 (Figure~\ref{modelfit}).
\\[4pt]
The hard state power spectra for Cygnus~X-1 and MAXI J1820+070 show broadband noise features in the frequency range $\sim$~5~mHz to  $\sim$~20~Hz, but their detailed spectral shapes are different (Figure~\ref{powspec}, Right panel). This is also reflected in the model fit parameters such as the variation of $\delta T_e$ as a function of frequency (Figure~\ref{modelfit}, Bottom Center panel). The detailed shapes of the variation is different for the two Cygnus~X-1 observations and for MAXI J1820+070. However, one sees that the attenuation factor $(\delta T_s/\delta T_e)$ for MAXI J1820+70 is remarkably similar to the January observation of Cygnus~X-1 for frequencies greater than $\sim$~2~Hz (Figure~\ref{modelfit}, Bottom Left panel). The time delay ($\tau_d$) between the variation of the soft inner disk temperature $(\delta T_s)$ and the inner region one $(\delta T_e)$ seems to have the same general frequency dependence for both the Cygnus~X-1 observations and the MAXI source, although the time delay for the MAXI source is lower at high frequencies (Figure~\ref{modelfit}, Bottom Right panel).
\\[4pt]
The temporal behavior of the system especially at high frequencies should depend on the geometry of the inner region which in turn would be characterized by physical parameters such as the inner disk radius and accretion rate. The different temporal behaviors seen for different Cygnus~X-1 observations maybe related to changes in these parameters. Differences between MAXI J1820+070 and Cygnus~X-1 could be additionally due to the systems having different black hole masses and that the latter being a transient while Cygnus~X-1 is a persistent source. Thus, it is interesting to note the qualitative similarity between the two sources especially the frequency dependence of the time delay $\tau_d$ which seems to be similar for both sources and for different Cygnus~X-1 observations, although the lower values seen at high frequency for MAXI J1820+070, may imply a lower black hole mass than Cygnus~X-1. The attenuation factor for the MAXI source is similar to the January observation at high frequencies implying that perhaps they were having similar geometries, but one should keep in mind the differences in the low frequency behavior and the overall PDS.
\section{Results and discussion}
\noindent
We report the results of spectro-timing analysis of MAXI J1820+070 as observed by SXT and LAXPC on-board {\em{AstroSat}} for the first time. We have analyzed $\sim$~93~ks of observed data corresponding to 15 individual satellite orbits. The broadband spectrum observed by SXT and LAXPC in the energy range 0.7$-$80.0~keV was well represented by a dominant thermal Comptonization component along with reflection and disk emission. The combined spectrum was modeled using a combination of \textit{tbabs, ireflect, nthcomp and diskbb} models. The PDS could be modeled with five Lorentzian components, a QPO with centroid frequency at 47.7~mHz, a weak oscillation at 109.4~mHz and three broadened noise humps spread over 0.004 to 30~Hz. This is the first confirmed report of detection of a QPO at 47.7~mHz in MAXI J1820+070 \citep{Mer18,Hon18} using \textit{AstroSat}. For a range of frequencies, LAXPC provides an unprecedented view of the energy dependent (3$-$80 keV) fractional rms and time-lags. \textit{AstroSat}'s capability of measuring the broadband time-averaged spectrum and the energy dependent temporal behavior of the system allows one to quantitatively fit both spectral and temporal data. 
\\[4pt]
The stochastic propagation model \citep{Maq19} is able to quantitatively fit the energy dependent temporal features. The fitting provides physical measurement of the time taken for a perturbation to travel from truncation radius to the inner regions of accretion disk. It also quantifies the normalized variation of temperatures due to inward propagation of perturbations. The results of modeling revealed that the perturbation time delay is of the order of 100~ms and is frequency dependent.
\section*{Acknowledgments}
\noindent
Authors would like to thank anonymous reviewer for the valued comments and suggestions that greatly improved this manuscript. This work began at IUCAA-APT Mini School on X-ray Astronomy Data Analysis held at Providence Women’s College, Kozhikode. SXT and LAXPC POC teams are thanked for their support
and timely release of data. This work has made use of software provided by HEASARC. SPM thanks CHRIST (Deemed to be University), Bengaluru for the `Research Assistant’ position through MRPDSC-1721. The authors (SBG $\&$ BSG) acknowledge Centre for Research, CHRIST (Deemed to be University), Bengaluru for the research funding (MRPDSC-1721). One of the authors (SBG) thanks the Inter-University Centre for Astronomy and Astrophysics(IUCAA), Pune for Visiting Associateship.

\begin{thebibliography}{}
\expandafter\ifx\csname natexlab\endcsname\relax\def\natexlab#1{#1}\fi
\providecommand{\url}[1]{\href{#1}{#1}}
\providecommand{\dodoi}[1]{doi:~\href{http://doi.org/#1}{\nolinkurl{#1}}}
\providecommand{\doeprint}[1]{\href{http://ascl.net/#1}{\nolinkurl{http://ascl.net/#1}}}
\providecommand{\doarXiv}[1]{\href{https://arxiv.org/abs/#1}{\nolinkurl{https://arxiv.org/abs/#1}}}

\bibitem[{{Abramowicz} {et~al.}(1995){Abramowicz}, {Chen}, \& {Taam}}]{Abr95}
{Abramowicz}, M.~A., {Chen}, X., \& {Taam}, R.~E. 1995, \apj, 452, 379

\bibitem[{{Ahmad} {et~al.}(2018){Ahmad}, {Misra}, {Iqbal}, {Maqbool}, \&
  {Hamid}}]{Ahm18}
{Ahmad}, N., {Misra}, R., {Iqbal}, N., {Maqbool}, B., \& {Hamid}, M. 2018, \na,
  58, 84

\bibitem[{{Antia} {et~al.}(2017){Antia}, {Yadav}, {Agrawal}, {Verdhan Chauhan},
  {Manchanda}, {Chitnis}, {Paul}, {Dedhia}, {Shah}, {Gujar}, {Katoch},
  {Kurhade}, {Madhwani}, {Manojkumar}, {Nikam}, {Pand ya}, {Parmar}, {Pawar},
  {Pahari}, {Misra}, {Navalgund}, {Pandiyan}, {Sharma}, \& {Subbarao}}]{Ant17}
{Antia}, H.~M., {Yadav}, J.~S., {Agrawal}, P.~C., {et~al.} 2017, \apjs, 231

\bibitem[{{Axelsson} \& {Done}(2018)}]{Axe18}
{Axelsson}, M., \& {Done}, C. 2018, \mnras, 480, 751

\bibitem[{{Baglio} {et~al.}(2018){Baglio}, {Russell}, \& {Lewis}}]{Bag18}
{Baglio}, M.~C., {Russell}, D.~M., \& {Lewis}, F. 2018, ATel, 11418

\bibitem[{{Belloni}(2010)}]{Bel10}
{Belloni}, T.~M. 2010, {States and Transitions in Black Hole Binaries}, Vol.
  794 (Springer-Verlag Berlin Heidelberg), 53

\bibitem[{{Bharali} {et~al.}(2019){Bharali}, {Chauhan}, \& {Boruah}}]{Bha19}
{Bharali}, P., {Chauhan}, J., \& {Boruah}, K. 2019, \mnras, 487, 5946

\bibitem[{{B{\"o}ttcher} \& {Liang}(1999)}]{Bot99}
{B{\"o}ttcher}, M., \& {Liang}, E.~P. 1999, \apjl, 511, L37

\bibitem[{{Chen}(1995)}]{Che95}
{Chen}, X. 1995, \mnras, 275, 641

\bibitem[{{Corral-Santana} {et~al.}(2016){Corral-Santana}, {Casares, J.},
  {Mu\~noz-Darias, T.}, {Bauer, F. E.}, {Mart\'{\i}nez-Pais, I. G.}, \&
  {Russell, D. M.}}]{Cor16}
{Corral-Santana}, J.~M., {Casares, J.}, {Mu\~noz-Darias, T.}, {et~al.} 2016,
  \aap, 587, A61

\bibitem[{{Cowperthwaite} \& {Reynolds}(2014)}]{Cow14}
{Cowperthwaite}, P.~S., \& {Reynolds}, C.~S. 2014, \apj, 791, 126

\bibitem[{{Esin} {et~al.}(1997){Esin}, {McClintock}, \& {Narayan}}]{Esi97}
{Esin}, A.~A., {McClintock}, J.~E., \& {Narayan}, R. 1997, \apj, 489, 865

\bibitem[{{Gandhi} {et~al.}(2019){Gandhi}, {Rao}, {Johnson}, {Paice}, \&
  {Maccarone}}]{Gan19}
{Gandhi}, P., {Rao}, A., {Johnson}, M. A.~C., {Paice}, J.~A., \& {Maccarone},
  T.~J. 2019, \mnras, 485, 2642

\bibitem[{{Gaskell}(2004)}]{Gas04}
{Gaskell}, C.~M. 2004, \apjl, 612, L21

\bibitem[{{Gleissner} {et~al.}(2004){Gleissner}, {Wilms}, {Pottschmidt},
  {Uttley}, {Nowak}, \& {Staubert}}]{Gle04}
{Gleissner}, T., {Wilms}, J., {Pottschmidt}, K., {et~al.} 2004, \aap, 414, 1091

\bibitem[{{Heil} {et~al.}(2012){Heil}, {Vaughan}, \& {Uttley}}]{Hei12}
{Heil}, L.~M., {Vaughan}, S., \& {Uttley}, P. 2012, \mnras, 422, 2620

\bibitem[{Hogg \& Reynolds(2016)}]{Hog16}
Hogg, J.~D., \& Reynolds, C.~S. 2016, \apj, 826, 40

\bibitem[{{Homan} {et~al.}(2018){Homan}, {Altamirano}, {Arzoumanian},
  {Buisson}, {Eikenberry}, {Fabian}, {Gendreau}, {Kara}, {Ludlam}, {Neilsen},
  {Ray}, {Remillard}, {Steiner}, {Uttley}, \& {Nicer Team}}]{Hon18}
{Homan}, J., {Altamirano}, D., {Arzoumanian}, Z., {et~al.} 2018, ATel, 11576

\bibitem[{{Ingram} \& {Done}(2011)}]{Ing11}
{Ingram}, A., \& {Done}, C. 2011, \mnras, 415, 2323

\bibitem[{{Ingram} \& {Done}(2012)}]{Ing12}
---. 2012, \mnras, 419, 2369

\bibitem[{{Ingram} \& {van der Klis}(2013)}]{Ing13}
{Ingram}, A., \& {van der Klis}, M. 2013, \mnras, 434, 1476

\bibitem[{{Kara} {et~al.}(2019){Kara}, {Steiner}, {Fabian}, {Cackett},
  {Uttley}, {Remillard}, {Gendreau}, {Arzoumanian}, {Altamirano}, {Eikenberry},
  {Enoto}, {Homan}, {Neilsen}, \& {Stevens}}]{Kar19}
{Kara}, E., {Steiner}, J.~F., {Fabian}, A.~C., {et~al.} 2019, \nat, 565, 198

\bibitem[{{Kawamuro} {et~al.}(2018){Kawamuro}, {Negoro}, {Yoneyama}, {Ueno},
  {Tomida}, {Ishikawa}, {Sugawara}, {Isobe}, {Shimomukai}, {Mihara},
  {Sugizaki}, {Nakahira}, {Iwakiri}, {Yatabe}, {Takao}, {Matsuoka}, {Kawai},
  {Sugita}, {Yoshii}, {Tachibana}, {Harita}, {Morita}, {Yoshida}, {Sakamoto},
  {Serino}, {Kawakubo}, {Kitaoka}, {Hashimoto}, {Tsunemi}, {Nakajima},
  {Kawase}, {Sakamaki}, {Maruyama}, {Ueda}, {Hori}, {Tanimoto}, {Oda},
  {Morita}, {Yamada}, {Tsuboi}, {Nakamura}, {Sasaki}, {Kawai}, {Sato},
  {Yamauchi}, {Hanyu}, {Hidaka}, {Yamaoka}, \& {Shidatsu}}]{Kaw18}
{Kawamuro}, T., {Negoro}, H., {Yoneyama}, T., {et~al.} 2018, ATel, 11399

\bibitem[{{Kotov} {et~al.}(2001){Kotov}, {Churazov}, \& {Gilfanov}}]{Kot01}
{Kotov}, O., {Churazov}, E., \& {Gilfanov}, M. 2001, \mnras, 327, 799

\bibitem[{{Kubota} {et~al.}(1998){Kubota}, {Tanaka}, {Makishima}, {Ueda},
  {Dotani}, {Inoue}, \& {Yamaoka}}]{Kub98}
{Kubota}, A., {Tanaka}, Y., {Makishima}, K., {et~al.} 1998, \pasj, 50, 667

\bibitem[{{Lyubarskii}(1997)}]{Lyu97}
{Lyubarskii}, Y.~E. 1997, \mnras, 292, 679

\bibitem[{{Magdziarz} \& {Zdziarski}(1995)}]{Mag95}
{Magdziarz}, P., \& {Zdziarski}, A.~A. 1995, \mnras, 273, 837

\bibitem[{{Mahmoud} \& {Done}(2018)}]{Mah18}
{Mahmoud}, R.~D., \& {Done}, C. 2018, \mnras, 480, 4040

\bibitem[{{Makishima} {et~al.}(1986){Makishima}, {Maejima}, {Mitsuda}, {Bradt},
  {Remillard}, {Tuohy}, {Hoshi}, \& {Nakagawa}}]{Mak86}
{Makishima}, K., {Maejima}, Y., {Mitsuda}, K., {et~al.} 1986, \apj, 308, 635

\bibitem[{{Maqbool} {et~al.}(2019){Maqbool}, {Mudambi}, {Misra}, {Yadav},
  {Gudennavar}, {Bubbly}, {Rao}, {Jogadand}, {Patil}, {Bhattacharyya}, \&
  {Singh}}]{Maq19}
{Maqbool}, B., {Mudambi}, S.~P., {Misra}, R., {et~al.} 2019, \mnras, 486, 2964

\bibitem[{{Mereminskiy} {et~al.}(2018){Mereminskiy}, {Grebenev}, {Molkov},
  {Zaznobin}, {Khorunzhev}, {Burenin}, \& {Eselevich}}]{Mer18}
{Mereminskiy}, I.~A., {Grebenev}, S.~A., {Molkov}, S.~V., {et~al.} 2018, ATel,
  11488

\bibitem[{{Misra}(2000)}]{Mis00}
{Misra}, R. 2000, \apjl, 529, L95

\bibitem[{{Misra} {et~al.}(2017){Misra}, {Yadav}, {Verdhan Chauhan}, {Agrawal},
  {Antia}, {Pahari}, {Chitnis}, {Dedhia}, {Katoch}, {Madhwani}, {Manchanda},
  {Paul}, \& {Shah}}]{Mis17}
{Misra}, R., {Yadav}, J.~S., {Verdhan Chauhan}, J., {et~al.} 2017, \apj, 835,
  195

\bibitem[{{Mitsuda} {et~al.}(1984){Mitsuda}, {Inoue}, {Koyama}, {Makishima},
  {Matsuoka}, {Ogawara}, {Shibazaki}, {Suzuki}, {Tanaka}, \& {Hirano}}]{Mit84}
{Mitsuda}, K., {Inoue}, H., {Koyama}, K., {et~al.} 1984, \pasj, 36, 741

\bibitem[{{Miyamoto} {et~al.}(1988){Miyamoto}, {Kitamoto}, {Mitsuda}, \&
  {Dotani}}]{Miy88}
{Miyamoto}, S., {Kitamoto}, S., {Mitsuda}, K., \& {Dotani}, T. 1988, \nat, 336,
  450

\bibitem[{{Narayan} \& {Yi}(1994)}]{Nar94}
{Narayan}, R., \& {Yi}, I. 1994, \apjl, 428, L13

\bibitem[{{Nowak} {et~al.}(1999){Nowak}, {Vaughan}, {Wilms}, {Dove}, \&
  {Begelman}}]{Now99}
{Nowak}, M.~A., {Vaughan}, B.~A., {Wilms}, J., {Dove}, J.~B., \& {Begelman},
  M.~C. 1999, \apj, 510, 874

\bibitem[{{Rapisarda} {et~al.}(2016){Rapisarda}, {Ingram}, {Kalamkar}, \& {van
  der Klis}}]{Rap16}
{Rapisarda}, S., {Ingram}, A., {Kalamkar}, M., \& {van der Klis}, M. 2016,
  \mnras, 462, 4078

\bibitem[{{Rapisarda} {et~al.}(2017){Rapisarda}, {Ingram}, \& {van der
  Klis}}]{Rap17}
{Rapisarda}, S., {Ingram}, A., \& {van der Klis}, M. 2017, \mnras, 472, 3821

\bibitem[{{Russell} {et~al.}(2019){Russell}, {Baglio}, \& {Lewis}}]{Rus19}
{Russell}, D.~M., {Baglio}, M.~C., \& {Lewis}, F. 2019, ATel, 12534

\bibitem[{{Shidatsu} {et~al.}(2019){Shidatsu}, {Nakahira}, {Murata}, {Adachi},
  {Kawai}, {Ueda}, \& {Negoro}}]{Shi19}
{Shidatsu}, M., {Nakahira}, S., {Murata}, K.~L., {et~al.} 2019, \apj, 874, 183

\bibitem[{{Spruit} {et~al.}(1987){Spruit}, {Matsuda}, {Inoue}, \&
  {Sawada}}]{Spr87}
{Spruit}, H.~C., {Matsuda}, T., {Inoue}, M., \& {Sawada}, K. 1987, \mnras, 229,
  517

\bibitem[{{Titarchuk} \& {Shaposhnikov}(2005)}]{Tit05}
{Titarchuk}, L., \& {Shaposhnikov}, N. 2005, \apj, 626, 298

\bibitem[{{Tucker} {et~al.}(2018){Tucker}, {Shappee}, {Holoien}, {Auchettl},
  {Strader}, {Stanek}, {Kochanek}, {Bahramian}, {ASAS-SN}, {Dong}, {Prieto},
  {Shields}, {Thompson}, {Beacom}, {Chomiuk}, {ATLAS}, {Denneau}, {Flewelling},
  {Heinze}, {Smith}, {Stalder}, {Tonry}, {Weiland}, {Rest}, {Huber}, {Rowan},
  \& {Dage}}]{Tuc18}
{Tucker}, M.~A., {Shappee}, B.~J., {Holoien}, T.~W.~S., {et~al.} 2018, \apjl,
  867, L9

\bibitem[{{Uttley} {et~al.}(2014){Uttley}, {Cackett}, {Fabian}, {Kara}, \&
  {Wilkins}}]{Utt14}
{Uttley}, P., {Cackett}, E.~M., {Fabian}, A.~C., {Kara}, E., \& {Wilkins},
  D.~R. 2014, \aapr, 22, 72

\bibitem[{{Uttley} \& {McHardy}(2001)}]{Utt01}
{Uttley}, P., \& {McHardy}, I.~M. 2001, \mnras, 323, L26

\bibitem[{{Uttley} {et~al.}(2018){Uttley}, {Gendreau}, {Markwardt},
  {Strohmayer}, {Bult}, {Arzoumanian}, {Pottschmidt}, {Ray}, {Remillard},
  {Pasham}, {Steiner}, {Neilsen}, {Homan}, {Miller}, {Iwakiri}, \&
  {Fabian}}]{Utt18}
{Uttley}, P., {Gendreau}, K., {Markwardt}, C., {et~al.} 2018, ATel, 11423

\bibitem[{{van der Klis}(1989)}]{Van89}
{van der Klis}, M. 1989, \araa, 27, 517

\bibitem[{{Wilms} {et~al.}(2000){Wilms}, {Allen}, \& {McCray}}]{Wil00}
{Wilms}, J., {Allen}, A., \& {McCray}, R. 2000, \apj, 542, 914

\bibitem[{{Zhang} {et~al.}(1997){Zhang}, {Cui}, {Harmon}, \&
  {Paciesas}}]{Zha97}
{Zhang}, S.~N., {Cui}, W., {Harmon}, B.~A., \& {Paciesas}, W.~S. 1997, in AIPC
  Series, Vol. 410, Proceedings of the Fourth Compton Symposium, ed. C.~D.
  {Dermer}, M.~S. {Strickman}, \& J.~D. {Kurfess}, 839--843

\bibitem[{{{\.Z}ycki} {et~al.}(1999){{\.Z}ycki}, {Done}, \& {Smith}}]{Zyc99}
{{\.Z}ycki}, P.~T., {Done}, C., \& {Smith}, D.~A. 1999, \mnras, 309, 561

\end{thebibliography}

\end{document}